# SOME CYLINDRICALLY SYMMETRIC VACUUM SOLUTIONS OF BRANS-DICKE SCALAR FIELDS IN ROBERTSON-WALKER UNIVERSE

KANGUJAM PRIYOKUMAR SINGH[1] and CHUNGKHAM GOKULCHANDRA SINGH [2].

[1]Department of Mathematical Sciences, Bodoland University, Kokrajhar.
BTC, Assam-783370
E-mail: pk_mathematics@yahoo.co.in

[2]Department of Mathematics, Presidency College, Motbung.

**Abstract.** The problem of cylindrically symmetric vacuum solutions of Brans-Dicke scalar fields has been studied. Exact solutions have been obtained for the vacuum B-D field equations for the cylindrically symmetric Einstein-Rosen metric. The solutions obtained in the present work are generalized solutions of the problem which has been studied by Rao et al. *(Annals of Physics, Vol. 87, 1974)*. The physical and kinematical behaviors of the solutions relevant to conformal space is also discussed in details, these solutions will be beneficial in solving the problems for investigating the different model of our universe.

## 1.Introduction

The development of cosmology the basic role played an idea of cosmological models together with an idea of astronomical tests [1]. The idea of Cosmological tests make cosmological models parts of astronomy which offers possibility of observationally determining the set of realistic parameters that can characterize variable models. While we can perform estimation of model parameters using a standard minimization procedure based on the likelihood method, may different scenarios are still in a good agreement with observational data of SNIa [2,3,4] as well as current measurement of cosmic microwave background anisotropies [5]. Also from the recent astronomical observations indicate that the universe is presently almost flat and undergoing a period of accelerated expansion. Basing on Einstein's general relativity all these observations can be explained by the hypothesis of a dark energy components in addition to cold dark matter (CDM) [6,7]. Also, the study of cosmological models in the frame work of general relativistic theory and in alternative theories of gravitation is active area of research for better understanding of the structure of the universe. In the general theory of relativity it has been found possible to introduce a scalar field without violating its essential features as discussed by the other authors in this field. The scalar field helps in the creation of matter in the cosmological theories. Brans-Dicke theory of gravitation is a well known competitor Einstein's theory [8,9]. The problem of Brans-Dicke scalar fields solutions interacting with gravitational field and several aspects of Brans-Dicke cosmology have been extensively investigated by many authors like [10-16].

Corresponding Author:
    Kangujam Priyokumar Singh, Department of Mathematical Sciences, Bodoland University, Kokrajhar.
    Kokrajhar- 783370, BTC, Assam, India.
    E-mail: pk_mathematics@yahoo.co.in

Here the problem of Brans-Dicke scalar fields vacuum solution for a cylindrically symmetric metric has been further studied and a more general solution has been obtained. The problem has been studied preserving all the necessary aspects already established in the original theory together with the Mach's principle. According to Dicke's opinion the inclusion of Mach's principle in the B-D theory, the problem requires a complete analysis of the situation by following a critical mathematical study, examining the implications of physics involved. The present work is a part of this investigation to make a substantial contribution of the B-D theory as compared to the other theories.

Rao et al. [17-19] have been taken up in the study of scalar fields in the Einstein's gravitational theory. They have obtained the exact solutions for the cylindrically symmetric Einstein-Rosen metric in the case of coupled source-free electromagnetic fields and zero-mass scalar fields. The solutions obtained have been compared with the results obtained by Gautreau [20]. Also, [21] discussed about Cylindrical Symmetric Brans-Dicke Fields II. On the other hand Dicke [9], Peters [22], Tabensky and Taub [23] have shown that the Brans-Dicke solutions are conformal to the solutions of Einstein's gravitational theory (either to vacuum solutions or zero-mass scalar fields).

Here, we have obtained exact solutions for the vacuum B-D field equations for the cylindrically symmetric Einstein-Rosen metric. The solutions obtained in the present work are generalized solutions of the problem which has been studied by [21]. These solutions of the vacuum field equations of the B-D scalar field for the cylindrically symmetric metric will be beneficial in solving the problems for investigating the different model of our universe.

.

## 1. Field Equations

The field equations of the B-D theory are

$$G^i_j = -\frac{k}{C^4}\frac{1}{\phi}T^i_j - \frac{\omega}{\phi^2}\left(\phi^i\phi_j - \frac{1}{2}\delta^i_j\phi^k\phi_k\right) - \frac{1}{\phi}\left(\phi^i_{;j} - \delta^i_j\phi^k_{;k}\right) \quad (1)$$

and

$$(3+2\omega)\phi^k_{;k} = \frac{k}{C^4}T, \quad (2)$$

where $G^i_j = R^i_j - \frac{1}{2}\delta^i_j R$ is the Einstein-tensor, $T^i_j$ is the stress-energy tensor, $T = T^k_k$, $\phi_i = \frac{\partial\phi}{\partial x}$ and $\omega$ is the coupling constant. In the following, we consider (1) and (2) in the case of a vacuum field, i.e., when $T^i_j = 0$.

In view of (2), we get the following three possibilities:

(i) $\phi^k_{;k} \neq 0, \quad \omega = -\frac{3}{2},$

(ii) $\phi^k_{;k} = 0, \quad \omega \neq -\frac{3}{2},$

(iii) $\phi^k_{;k} = 0, \quad \omega = -\frac{3}{2}.$

We observe that (ii) and (iii) lead to the same result and thus we consider the cases (i) and (ii) only. The field equations (1) and (2) for the case (i) reduce to only one equation, viz.,

$$G^i_j = \frac{3}{2\phi^2}\left(\phi^i \phi_j - \frac{1}{2}\delta^i_j \phi^k \phi_k\right) - \frac{1}{\phi}\left(\phi^i_{;j} - \delta^i_j \phi^k_{;k}\right). \tag{3}$$

For the case (ii), the B-D field equations are given by

$$G^i_j = -\frac{\omega}{\phi^2}\left(\phi^i \phi_j - \frac{1}{2}\delta^i_j \phi^k \phi_k\right) - \frac{1}{\phi} \phi^i_{;j} \tag{4}$$

and

$$\phi^k_{;k} = g^{ij} \phi_{;ij}. \tag{5}$$

For the present problem the metric considered is

$$ds^2 = e^{2\alpha - 2\beta}(dt^2 - d\rho^2) - \rho^2 e^{-2\beta} d\varphi^2 - e^{2\beta} dz^2 \tag{6}$$

where $\alpha$, $\beta$ are functions of $\rho$ and t only and $\rho$, $\varphi$, z, t correspond to $x^1$, $x^2$, $x^3$, $x^4$ co-ordinates respectively. The axial symmetry assumed impulse that $\phi$ shares the same symmetry as $\alpha$ and $\beta$. As a consequence of which we note that

$$\phi_2 = \phi_3 = 0. \tag{7}$$

Here and in what follows the lower suffixes 1, 2, 3, 4 after an unknown function will mean the partial differentiation with respect to $\rho$, $\varphi$, z and t respectively.

The surviving field equations for the metric (6) are

$$G^1_1 = -e^{2\beta - 2\alpha}\left(\beta_1^2 + \beta_4^2 - \frac{\alpha_1}{\rho}\right), \tag{8}$$

$$G_2^2 = e^{2\beta - 2\alpha}\left(\alpha_{11} - \alpha_{44} + \beta_1^2 - \beta_4^2\right), \tag{9}$$

$$G_3^3 = e^{2\beta - 2\alpha}\left(2\beta_{44} - 2\beta_{11} - \frac{2\beta_1}{\rho} + \alpha_{11} - \alpha_{44} - \beta_4^2 + \beta_1^2\right), \tag{10}$$

$$G_4^4 = e^{2\beta - 2\alpha}\left(\beta_1^2 + \beta_4^2 - \frac{\alpha_1}{\rho}\right), \tag{11}$$

$$G_1^4 = -G_4^1 = e^{2\beta - 2\alpha}\left(2\beta_1\beta_4 - \frac{\alpha_4}{\rho}\right). \tag{12}$$

## 2. Solution of the Field Equations

We now consider the solutions of the field equations for the cases mentioned in the previous section.

The field Equations (3) for the case $\phi^k_{;k} \neq 0$, $\omega = -\frac{3}{2}$ reduce to

$$G_1^1 \equiv \beta_1^2 + \beta_4^2 - \frac{\alpha_1}{\rho} = \frac{3}{4}\frac{1}{\phi^2}\left(\phi_1^2 + \phi_4^2\right) + \frac{1}{\phi}\left\{(\alpha_1 - \beta_1)\phi_1 + (\alpha_4 - \beta_4)\phi_4 + \frac{\phi_1}{\rho} - \phi_{44}\right\} \tag{13}$$

$$G_2^2 \equiv \alpha_{11} - \alpha_{44} + \beta_1^2 - \beta_4^2 = -\frac{3}{4}\frac{1}{\phi^2}(\phi_4^2 - \phi_1^2) - \frac{1}{\phi}(\beta_1\phi_1 - \beta_4\phi_4 + \phi_{11} - \phi_{44}), \tag{14}$$

$$G_3^3 \equiv 2\beta_{44} - 2\beta_{11} - \frac{2\beta_1}{\rho} + \alpha_{11} - \alpha_{44} + \beta_1^2 - \beta_4^2$$
$$= -\frac{3}{4}\frac{1}{\phi^2}\left(\phi_4^2 - \phi_1^2\right) - \frac{1}{\phi}\left(\beta_4\phi_4 - \beta_1\phi_1 - \phi_{44} + \phi_{11} + \frac{\phi_1}{\rho}\right), \tag{15}$$

$$G_4^4 \equiv \beta_1^2 + \beta_4^2 - \frac{\alpha_1}{\rho} = \frac{3}{4}\cdot\frac{1}{\phi^2}\left(\phi_1^2 + \phi_4^2\right) - \frac{1}{\phi}\left\{\phi_{11} + \frac{\phi_1}{\rho} - (\alpha_1 - \beta_1)\phi - (\alpha_4 - \beta_4)\phi_4\right\}, \tag{16}$$

$$G_4^1 = -G_1^4 \equiv 2\beta_1\beta_4 - \frac{\alpha_4}{\rho} = \frac{3}{2}\frac{1}{\phi^2}\phi_1\phi_4 + \frac{1}{\phi}\left\{(\alpha_4 - \beta_4)\phi_1 + (\alpha_1 - \beta_1)\phi_4 - \phi_{14}\right\}. \tag{17}$$

From Equations (13) and (16), we obtain

$$\phi_{11} - \phi_{44} + \frac{2\phi_1}{\rho} = 0. \tag{18}$$

It can be verified that Equation (14) can be derived from Equations (13), (15), (16) and (17) as the integrability condition on $\alpha$. Hence, we need to solve Equations (13), (15), (17) and (18) only for $\alpha$, $\beta$ and $\phi$.

When $\alpha$, $\beta$, $\phi$ are functions of $\rho$ only, the field Equations (13), (14), (15) and (16) reduce to the following equations

$$\beta_1^2 - \frac{\alpha_1}{\rho} = \frac{3}{4}\left(\frac{\phi_1}{\phi}\right)^2 + \frac{1}{\phi}\left\{(\alpha_1 - \beta_1)\phi_1 + \frac{\phi_1}{\rho}\right\}, \qquad (19)$$

$$\alpha_{11} + \beta_1^2 = \frac{3}{4}\left(\frac{\phi_1}{\phi}\right)^2 - \frac{1}{\phi}(\beta_1 \phi_1 + \phi_{11}), \qquad (20)$$

$$\alpha_{11} - 2\beta_{11} - \frac{2\beta_1}{\rho} + \beta_1^2 = \frac{3}{4}\left(\frac{\phi_1}{\phi}\right)^2 + \frac{1}{\phi}\left(\beta_1 \phi_1 - \phi_{11} - \frac{\phi_1}{\rho}\right), \qquad (21)$$

and

$$\beta_1^2 - \frac{\alpha_1}{\rho} = \frac{3}{4}\left(\frac{\phi_1}{\phi}\right)^2 - \frac{1}{\phi}\left\{\phi_{11} + \frac{\phi_1}{\rho} - (\alpha_1 - \beta_1)\phi_1\right\}. \qquad (22)$$

From Equations (19) and (22), we obtain, on comparing

$$\phi_{11} + \frac{2\phi_1}{\rho} = 0. \qquad (23)$$

From Equation (23), we obtain

$$\phi = -\frac{a}{\rho} + b, \qquad (24)$$

where a and b are arbitrary constants.

From Equations (19), (20), (21), (22) and using equation (24), we obtain

$$\beta = \frac{1}{2}\log\{\rho(b\rho - a)^{-1}\} + \frac{c}{b}\log(b\rho - a), \qquad (25)$$

where c is also an arbitrary constant.

Using equations (24) and (25) in (19), we obtain

$$\alpha = \log\rho + \frac{c^2 - b^2}{b^2}\log(b\rho - a). \qquad (26)$$

The metric (6), thus reduces to

$$ds^2 = \rho(b\rho - a)^{\frac{2c^2 - b^2 - 2bc}{b^2}}(dt^2 - d\rho^2) - \rho(b\rho - a)^{\frac{b - 2c}{b}}d\varphi^2 - \rho(b\rho - a)^{\frac{2c - b}{b}}dz^2. \quad (27)$$

The field Equations (4) and (5) for the case $\phi^k_{;k} = 0$, $\omega \neq -\frac{3}{2}$ reduce to

$$G_1^1 \equiv \beta_1^2 + \beta_4^2 - \frac{\alpha_1}{\rho} = -\frac{\omega}{2}\frac{1}{\phi^2}(\phi_1^2 + \phi_4^2) + \frac{1}{\phi}\{(\alpha_1 - \beta_1)\phi_1 + (\alpha_4 - \beta_4)\phi_4 - \phi_{11}\}, \qquad (28)$$

$$G_2^2 \equiv \alpha_{11} - \alpha_{44} + \beta_1^2 - \beta_4^2 = \frac{\omega}{2}\frac{1}{\phi^2}(\phi_4^2 - \phi_1^2) - \frac{1}{\phi}\left\{\left(\beta_1 - \frac{1}{\rho}\right)\phi_1 - \phi_4 \beta_4\right\}, \qquad (29)$$

$$G_3^3 \equiv 2\beta_{44} - 2\beta_{11} - \frac{2\beta_1}{\rho} + \alpha_{11} - \alpha_{44} + \beta_1^2 - \beta_4^2$$
$$= \frac{\omega}{2}\frac{1}{\phi^2}\left(\phi_4^2 - \phi_1^2\right) - \frac{1}{\phi}\left(\beta_4\phi_4 - \beta_1\phi_1\right), \tag{30}$$

$$G_4^4 \equiv \beta_1^2 + \beta_4^2 - \frac{\alpha_1}{\rho} = -\frac{\omega}{2}\frac{1}{\phi^2}\left(\phi_1^2 + \phi_4^2\right) - \frac{1}{\phi}\left\{\phi_{44} - (\alpha_1 - \beta_1)\phi - (\alpha_4 - \beta_4)\phi_4\right\}, \tag{31}$$

$$G_4^1 = -G_1^4 \equiv 2\beta_1\beta_4 - \frac{\alpha_4}{\rho}$$
$$= -\frac{\omega}{\phi^2}\phi_1\phi_4 + \frac{1}{\phi}\left\{(\alpha_4 - \beta_4)\phi_1 + (\alpha_1 - \beta_1)\phi_4 - \phi_{14}\right\} \tag{32}$$

and

$$\phi_{11} - \phi_{44} + \frac{\phi_1}{\rho} = 0. \tag{33}$$

From Equations (28) and (31), we obtain

$$\phi_{44} - \phi_{11} = 0. \tag{34}$$

From Equations (33) and (34), we obtain

$$\phi = a_1 t + b_1, \tag{35}$$

where $a_1$ and $b_1$ are arbitrary constants.

It can be verified that Equation (29) may be derived from Equations (28), (30), (31), (32) and (33) as an integrability condition on $\alpha$. Hence, we need to solve Equations (28), (30), (31) and (32) for the unknowns $\alpha$ and $\beta$.

The field Equations (28), (30), (31) and (32) reduce to the following

$$\alpha_1\left(\frac{1}{\rho^2} - \frac{\phi_4^2}{\phi^2}\right) = \frac{1}{\rho}\left(\beta_1^2 + \beta_4^2\right) - 2\beta_1\beta_4\frac{\phi_4}{\phi} - \beta_1\frac{\phi_4^2}{\phi^2} + \beta_4\frac{\phi_4}{\phi\rho} + \frac{\omega\phi_4^2}{2\rho\phi^2}, \tag{36}$$

$$\alpha_4\left(\frac{1}{\rho^2} - \frac{\phi_4^2}{\phi^2}\right) = -\frac{\phi_4}{\phi}\left(\beta_1^2 + \beta_4^2\right) + 2\beta_1\beta_4\cdot\frac{1}{\rho} + \beta_1\cdot\frac{\phi_4}{\phi\rho} - \beta_4\cdot\frac{\phi_4^2}{\phi^2} - \frac{\omega}{2}\frac{\phi_4^3}{\phi^3} \tag{37}$$

and

$$\beta_{44} - \beta_{11} - \frac{\beta_1}{\rho} = -\frac{\phi_4}{\phi}\beta_4. \tag{38}$$

From Equation (38), we obtain

$$\beta = \frac{h(a_1 t + b_1)^2}{4a_1^2} + \frac{\ell}{a_1}\log(a_1 t + b_1) + \frac{h\rho^2}{4} + n\log\rho, \tag{39}$$

where h is a separation constant, and $\ell$, n are arbitrary constants.

Substituting the values of φ and β in (36), we obtain

$$\alpha = -\frac{n^2}{2}\log\left\{\frac{1}{\rho^2} - \frac{a_1^2}{(a_1 t + b_1)^2}\right\} + \frac{1}{2a_1^2}\left\{\ell(2a_1 n - \ell) + a_1(a_1 n - \ell) - \frac{\omega a_1^2}{2}\right\} \times \log\left\{(a_1 t + b_1)^2 - a_1^2 \rho^2\right\}$$

$$+ \left\{\frac{h^2}{4} + \frac{a_1 h\ell}{(a_1 t + b_1)^2} + \frac{h a_1^2}{2(a_1 t + b_1)^2}\right\} \frac{(a_1 t + b_1)^2 \rho^2}{2a_1^2} \quad (40)$$

Similarly, from Equation (37), we obtain

$$\alpha = -\frac{1}{2a_1}(a_1 n^2 - 2n\ell - n a_1) \log\left\{(a_1 t + b_1)^2 - a_1^2 \rho^2\right\} -$$

$$- \frac{1}{2a_1^3}\left(a_1 \ell^2 + \ell a_1^2 + \frac{\omega a_1^3}{2}\right) \log\left\{\frac{1}{\rho^2} - \frac{a_1^2}{(a_1 t + b_1)^2}\right\} + \left(\frac{h^2}{4a_1} + \frac{nh}{a_1 \rho^2}\right) \frac{\rho^2 (a_1 t + b_1)^2}{2a_1} . \quad (41)$$

Comparing equations (40) and (41), we obtain

$$\alpha = -\frac{n^2}{2}\log\left\{\frac{1}{\rho^2} - \frac{a_1^2}{(a_1 t + b_1)^2}\right\} + \frac{1}{2a}(2n\ell - a_1 n^2 + a_1 n) \log\left\{(a_1 t + b_1)^2 - a_1^2 \rho^2\right\} +$$

$$+ \frac{h^2 \rho^2 (a_1 t + b_1)^2}{8a_1^2} + \frac{h\ell\rho^2}{2a_1} + \frac{h\rho^2}{4} + \frac{hn(a_1 t + b_1)^2}{2a^2} . \quad (42)$$

subject to the relation between constants

$$n^2 = \frac{\ell}{a_1}\left(\frac{\ell}{a_1} + 1\right) + \frac{\omega}{2} . \quad (43)$$

Thus the metric (6) reduces to

$$ds^2 = \frac{\left\{(a_1 t + b_1)^2 - a_1^2 \rho^2\right\}^{\frac{1}{a_1}\left(2n\ell - a_1 n^2 + a_1 n\right)} \cdot e^{\frac{(a_1 t + b_1)^2}{a_1^2}\left[\left\{\frac{h}{4} + \frac{a_1 \ell}{(a_1 t + b_1)^2}\right\} h\rho^2 + \left(n - \frac{1}{2}\right) h\right]}}{\left\{\frac{1}{\rho^2} - \frac{a_1^2}{(a_1 t + b_1)^2}\right\}^{n^2} (a_1 t + b_1)^{\frac{2\ell}{a_1}} \rho^{2n}} \times (dt^2 - d\rho^2)$$

$$-\rho^{2-2n} \cdot e^{-h\left\{\frac{(a_1 t + b_1)^2}{2a_1^2} + \frac{\rho^2}{2}\right\}} (a_1 t + b_1)^{\frac{2\ell}{a_1}} d\varphi^2 - \rho^{2n} e^{h\left\{\frac{(a_1 t + b_1)^2}{2a_1^2} + \frac{\rho^2}{2}\right\}} (a_1 t + b_1)^{\frac{2\ell}{a_1}} dz^2. \quad (44)$$

## 3. Discussion of the solutions

We can here note that the Brans-Dicke vacuum scalar fields are conformal to zero-mass scalar fields of the gravitational theory as shown by Dicke [9] and Tabensky and Taub [23]. We can also verify that when the coupling constant $\omega = -\frac{3}{2}$, the B-D field is conformal to vacuum solutions of the gravitational theory and not to the zero-mass scalar field [9,23]. For the case $\omega \neq -\frac{3}{2}$, the corresponding relations between the vacuum B-D field and zero-mass scalar field in Einstein's theory have been given by Tabensky and Taub [23] These relations are given by

$$e^{\frac{\sqrt{2}\,V}{(\omega + 3/2)^{1/2}}} = \phi_{B-D} \tag{45}$$

and

$$g_{\mu\nu} = \phi_{B-D} \cdot g_{B-D\mu\nu} \tag{46}$$

where $\phi_{B-D}$ and $g_{\mu\nu}$ are the quantities occurring in the B-D theory. In the present work, we have considered both the above cases (i.e. $\omega = -\frac{3}{2}$ and $\omega \neq -\frac{3}{2}$) separately and obtained the solutions for each case. We also would like to note further that the solutions obtained for the case $\omega = -\frac{3}{2}$ (being conformal to vacuum solutions) have no corresponding known solutions from which they can be generated by applying a conformal transformation. In the case of $\omega \neq -\frac{3}{2}$, the solutions corresponding to the metric (44) in the conformal space, using the relations (45) and (46) are given by

$$\begin{aligned}
g_{11} &= -g_{44} \\
&= -\tau^H \cdot \rho^F \cdot \exp\left\{\frac{\tau^2}{a_1^2}\left(\frac{h}{4} + \frac{a_1 \ell}{\tau^2}\right) h \rho^2 + \left(n - \frac{1}{2}\right) h\right\} \bigg/ \left\{\frac{1}{\rho^2} - \frac{a_1^2}{\tau^2}\right\}^E \cdot \\
g_{22} &= -\rho^{2(1-n)} \cdot \exp\left\{-h\left(\frac{\tau^2}{2a_1^2} + \frac{\rho^2}{2}\right)\right\} \cdot \tau^{1 - \frac{2\ell}{a_1}} \\
g_{33} &= -\rho^{2n} \cdot \exp\left\{h\left(\frac{\tau^2}{2a_1^2} + \frac{\rho^2}{2}\right)\right\} \cdot \tau^{1 + \frac{2\ell}{a_1}}
\end{aligned} \tag{47}$$

and

$$V = \frac{1}{\sqrt{2}}\left(\omega + \tfrac{3}{2}\right)^{\frac{1}{2}} \log \tau,$$

where we denote

$$\tau = a_1 t + b_1$$

$$H = \frac{2}{a_1}\left(2n\ell - a_1 n^2 + a_1 n - \ell + \frac{a_1}{2}\right)$$

$$F = \frac{2}{a_1}\left(2n\ell - a_1 n^2\right)$$

$$E = 2n^2 - \frac{2\ell n}{a_1} - n$$

These are the zero-mass solutions of the most general cylindrically symmetric metric of Marder [24], given by (Vide appendix – II at the end of this chapter)

$$ds^2 = e^{2\alpha - 2\beta}\left(dt^2 - d\rho^2\right) - \rho^2 e^{-2\beta} d\varphi^2 - e^{2\beta + 2\gamma} dz^2 , \qquad (47)$$

and not of the Einstein-Rosen metric (6). The solutions thus obtained by conformal transformations either for the case $\omega = -\frac{3}{2}$ or, $\omega \neq -\frac{3}{2}$, of which one case will go over to the vacuum solution and the other case to the zero-mass scalar field of Einstein's gravitational theory, are to be studied further. We may here note that from the observational point of view the value of the coupling constant $\omega$ is approximately equal to 6. However the case $\omega = -\frac{3}{2}$ has given some results which of physical interest because of their origin being generated from Einstein's vacuum solutions.

## APPENDIX - I

For the metric (6), the non-vanishing Christoffel symbols are

$$\Gamma^1_{11} = \Gamma^1_{44} = \Gamma^4_{14} = \alpha_1 - \beta_1 ,$$

$$\Gamma^4_{11} = \Gamma^1_{14} = \Gamma^4_{44} = \alpha_4 - \beta_4 ,$$

$$\Gamma^2_{12} = \frac{1}{\rho} - \beta_1 = -e^{-2\alpha} \cdot \frac{1}{\rho^2} \Gamma^1_{22} ,$$

$$\Gamma^2_{42} = -\beta_4 = \frac{e^{2\alpha}}{\rho^2} \Gamma^4_{22} ,$$

$$\Gamma^3_{13} = \beta_1 = -e^{2\alpha - 4\beta} \Gamma^1_{33}$$

and

$$\Gamma^3_{43} = \beta_4 = e^{2\alpha - 4\beta} \Gamma^4_{33}$$

The non-zero components of the curvature tensor are given by

$$R_{1212} = -\rho^2 e^{-2\beta}\left(\beta_{11} - \alpha_1\beta_1 - \alpha_4\beta_4 + \beta_4^2 + \frac{\alpha_1}{\rho} + \frac{\beta_1}{\rho}\right),$$

$$R_{1313} = e^{2\beta}\left(\beta_{11} - \alpha_1\beta_1 + 2\beta_1^2 - \alpha_4\beta_4 + \beta_4^2\right),$$

$$R_{2323} = \rho^2 e^{2\beta - 2\alpha}\left(\beta_4^2 - \beta_1^2 + \frac{\beta_1}{\rho}\right),$$

$$R_{1414} = e^{2\alpha - 2\beta}\left(\alpha_{44} - \alpha_{11} + \beta_{11} - \beta_{44}\right),$$

$$R_{2424} = -\rho^2 e^{-2\beta}\left(\beta_{44} - \alpha_4\beta_4 + \beta_1^2 - \alpha_1\beta_1 + \frac{\alpha_1}{\rho} - \frac{\beta_1}{\rho}\right),$$

$$R_{3434} = e^{2\beta}\left(\beta_{44} - \alpha_4\beta_4 + 2\beta_4^2 - \alpha_1\beta_1 + \beta_1^2\right),$$

$$R_{1224} = \rho^2 e^{-2\beta}\left(\beta_{14} + \beta_1\beta_4 - \beta_1\alpha_4 - \alpha_1\beta_4 + \frac{\alpha_4}{\rho}\right)$$

and

$$R_{1334} = -e^{2\beta}\left(\beta_{14} + 3\beta_1\beta_4 - \beta_1\alpha_4 - \alpha_1\beta_4\right).$$

### APPENDIX – II

The Einstein gravitational field equations for Marder's metric (48) for zero-mass scalar field as energy-momentum tensor are as follows :

$$G_1^1 \equiv -\gamma_{44} + \alpha_1\gamma_1 + \alpha_4\gamma_4 - 2\beta_1\gamma_1 - 2\beta_4\gamma_4 - \beta_1^2 - \beta_4^2 - \gamma_4^2 + \frac{\alpha_1}{\rho} + \frac{\gamma_1}{\rho} = \frac{k}{8\pi}\left(V_1^2 + V_4^2\right),$$

$$G_2^2 \equiv \alpha_{11} - \alpha_{44} + \gamma_{11} - \gamma_{44} + 2\beta_1\gamma_1 - 2\beta_4\gamma_4 + \beta_1^2 - \beta_4^2 + \gamma_1^2 - \gamma_4^2 = \frac{k}{8\pi}\left(V_4^2 - V_1^2\right),$$

$$G_3^3 \equiv \alpha_{11} - \alpha_{44} + 2\beta_{44} - 2\beta_{11} - \frac{2\beta_1}{\rho} + \beta_1^2 - \beta_4^2 = \frac{k}{8\pi}\left(V_4^2 - V_1^2\right),$$

$$G_4^4 \equiv \gamma_{11} - \alpha_1\gamma_1 - \alpha_4\gamma_4 + 2\beta_4\gamma_4 + 2\beta_1\gamma_1 + \beta_4^2 - \frac{\alpha_1}{\rho} + \beta_1^2 + \gamma_1^2 + \frac{\gamma_1}{\rho} = -\frac{k}{8\pi}\left(V_1^2 + V_4^2\right),$$

$$G_1^4 \equiv -G_4^1 = \gamma_{14} + \gamma_1\gamma_4 + 2\beta_1\beta_4 - \alpha_1\gamma_4 - \alpha_4\gamma_1 + 2\beta_1\gamma_4 + 2\beta_4\gamma_1 - \frac{\alpha_4}{\rho} = \frac{k}{4\pi}V_1V_4$$

and

$$g^{ij}V_{;ij} \equiv 0 = V_{44} - V_{11} - \frac{V_1}{\rho} - V_1\gamma_1 + V_4\gamma_4.$$